# MODIFIED BERNOULLI EQUATION FOR USE WITH COMBINED ELECTRO-OSMOTIC AND PRESSURE-DRIVEN MICROFLOWS

**T. M. Adams and A. Raghunandan**
**Department of Mechanical Engineering**
**Rose-Hulman Institute of Technology**
**Terre Haute, IN, USA**
**adams1@rose-hulman.edu**

**ABSTRACT**

In this paper we present electro-osmotic (EO) flow within a more traditional fluid mechanics framework. Specifically, the modified Bernoulli equation (viz. the energy equation, the mechanical energy equation, the pipe flow equation, etc.) is shown to be applicable to EO flows if an electrical potential energy term is also included. The form of the loss term in the modified Bernoulli equation is unaffected by the presence of an electric field; i.e., the loss term still represents the effect of wall shear stress, which can be represented via a friction factor. We show that that the friction factor for pure EO flow (no applied pressure gradient) varies inversely with the Reynolds number based on the Debeye length of the electric double layer. Expressions for friction factor for combined laminar pressure-driven and EO flow are also given. These are shown to be functions of Reynolds number and geometry, as well as the relative strength of the applied electric field to the applied pressure gradient.

## INTRODUCTION

Microfluidics continues to play one of the largest roles in the development of some of the most innovative research applications. This is perhaps most noticeable within chemistry and the life sciences where the amounts of materials and the time required for analyses have been minimized considerably by making use of microfluidic systems. Many of these micro-total analysis system (μ-TAS) or "lab-on-a-chip" systems involve capillary electrophoretic separations on glass microchips in which the flow is accomplished via electro-osmosis (EO). Pressure-driven microfluidic devices are also common, and some applications, such as electro-kinetic pumping [1], involve both EO and pressure driven flow.

Though there have been several investigations of the fluid mechanics of EO flow in microchannels [2]-[4], as well as of combined EO and pressure driven flow [5]-[6], taken as a whole there seems to be a lack of cohesion in the presented results, and few general conclusions seem to have been drawn. This is due in part to the fact that the fluid mechanics required to understand microfluidic devices often involve unique solutions tailored to the application at hand. Furthermore, much of the physics of EO flow was investigated well before the advent microfluidics. The development of much of microfluidics, therefore, has developed somewhat independently of traditional engineering fluid mechanics. This paper aims to make some of the flow solutions encountered in microfluidics more accessible to a larger audience by framing the results in a more traditional form. Namely, combined electro-osmotic and pressure-driven flow is cast in the form of the modified Bernoulli equation and the friction factor correlations for such flows are derived.

## MODIFIED BERNOULLI EQUATION

For the adiabatic and reversible flow of an incompressible fluid with no additions or extractions of energy, the Bernoulli equation can be applied between any two points along a fluid stream to give

$$P_1 + \rho \frac{V_1^2}{2} + \rho g z_1 = P_2 + \rho \frac{V_2^2}{2} + \rho g z_2. \qquad (1)$$

The Bernoulli equation is a statement that the mechanical energy of the flowing fluid remains constant, though it can be distributed differently among pressure, kinetic energy and gravitational potential energy forms at different points in the flow direction. In cases in which energy can be added or extracted from the fluid stream via fluid machines, or in which dissipative effects such as wall friction are present (Fig. 1), the Bernoulli equation is modified:

$$P_1 + \alpha_1 \rho \frac{V_1^2}{2} + \rho g z_1 + \rho g h_{in}$$
$$= P_2 + \alpha_2 \rho \frac{V_2^2}{2} + \rho g z_2 + \rho g h_{out} + \rho g h_{loss} \quad (2)$$

Equation (2) is referred to as the modified Bernoulli equation, the extended Bernoulli equation, the mechanical energy equation, and the pipe flow equation, as well as by other names. It is one of the most extensively used and well known equations in fluid mechanics, most likely familiar to the reader from his/her first course in elementary fluid mechanics. The form given in (2) also includes the kinetic energy correction factor, $\alpha$, which accounts for the possibility of non-uniform velocity profiles, such as in fully developed laminar flow.

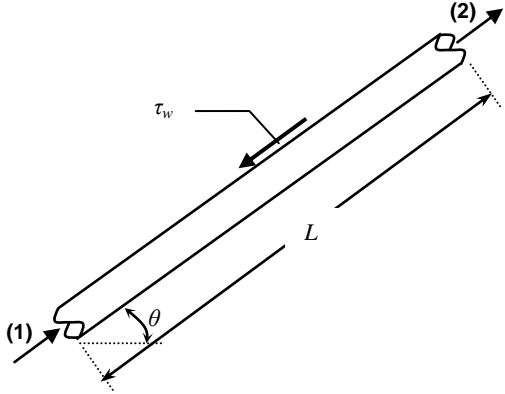

Fig 1: Schematic fluid flow system with wall friction

Unlike the pressure-driven flows for which the modified Bernoulli equation is usually employed, in electro-osmotic (EO) flow an applied electric field in the flow direction is responsible for generating the flow. For certain fluid/solid combinations, complicated surface chemical interactions result in a thin non-zero layer of electric charge next to the solid surface, called the electric double layer (EDL). An applied field causes these "charge plates" to move in the direction of decreasing electric potential, viscously dragging the fluid it bounds. This causes the almost completely flat velocity profile seen in EO flow, one of the traits that makes it so attractive for chemical and biological separations and detection in microfluidic devices. Figure 2 gives a schematic of this simplified view of EO flow.

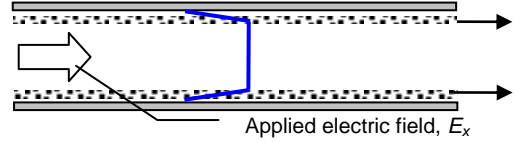

Fig. 2: Electro-osmotic flow

The velocity distribution for EO flow is well known and good treatments for its development can be found elsewhere [7]-[8]. The major result of concern to us here is the exponential nature of the flow direction velocity as a function of the coordinate perpendicular to the solid surface, $y$:

$$u(y) = \frac{\varepsilon \zeta E_x}{\eta}\left(e^{-(y/\lambda_D)} - 1\right), \quad (3)$$

where $\zeta$ is the effective wall electric potential (called the zeta potential), $\varepsilon$ is the electrical permittivity of the fluid, $\eta$ is viscosity, and $\lambda_D$ is the Debye length. The Debye length represents the exponential decay distance for the wall electric potential and is on the order of 1-100 nm in most cases. Given that the Debye length is much smaller than the widths of most microfluidic channels, the flow velocity reaches its constant average value a short distance from the channel wall, which is found by letting $y\to\infty$ in (3).

$$u_{EO,avg} = \frac{\varepsilon \zeta E_x}{\eta}. \quad (4)$$

One of the goals of the present work is to show that electro-osmotic flow can also be captured by the modified Bernoulli equation. This is accomplished by introducing electric potential terms to (2),

$$P_1 + \alpha_1 \rho \frac{V_1^2}{2} + \rho g z_1 + \rho_e \varphi_1 + \rho g h_{in}$$
$$= P_2 + \alpha_2 \rho \frac{V_2^2}{2} + \rho g z_2 + \rho_e \varphi_2 + \rho g h_{out} + \rho g h_{loss} \quad (5)$$

where $\rho_e$ is the charge density of the electrolyte and $\varphi$ is the electrical potential (voltage) at any point along the flow direction. The $\rho_e \varphi$ terms represent the *electrical* potential energy of the fluid. In the event that no applied electric field exists, $\varphi_2 - \varphi_1 = 0$ and the original pipe flow equation is retrieved.

In order to make use of (5) the form of the loss term $h_{loss}$ must be explored. Assuming a constant electric field applied in the flow direction and no fluid machines adding or extracting energy from the fluid stream, a macroscopic momentum balance for the system in Fig. 1 yields

$$P_1 - P_2 = \rho g L \sin\theta + \rho_e(\varphi_2 - \varphi_1) + \frac{\tau_w \mathcal{P} L}{A_c}, \quad (6)$$

where $\mathcal{P}$ is the wetted perimeter and $A_c$ is the cross sectional flow area of the flow channel. Eliminating pressure from (5) and (6) gives

$$\rho g h_{loss} = \frac{\tau_w \mathcal{P} L}{A_c}. \quad (7)$$

Thus the loss term is still a manifestation of wall shear stress just as it is in purely pressure driven flow. It therefore makes sense to speak of a D'Arcy friction factor even in the presence of an electric field:

$$f \equiv \frac{4\tau_w}{\frac{1}{2}\rho u_{avg}^2}, \quad (8)$$

the viscous shear stress at the wall being calculated from

$$\tau_w = \eta \left(\frac{du}{dy}\right)_{y=0}. \quad (9)$$

In the case of pure EO flow (no applied pressure gradient) using the velocity distribution of (3) in (9) gives

$$\tau_w = \frac{\varepsilon \zeta E_x}{\lambda_D} = u_{EO,avg} \frac{\eta}{\lambda_D}. \quad (10)$$

Substituting (10) into (8) yields the friction factor for EO flow:

$$f = \frac{8}{\frac{\rho u_{avg} \lambda_D}{\eta}} = \frac{8}{Re_{\lambda_D}}. \quad (11)$$

This expression for $f$ is independent of geometry, since the velocity distribution of (3) depends on the Debeye length, $\lambda_D$, being small compared to the length scale of the channel. If one prefers friction factor expressed in terms of hydraulic diameter (11) becomes

$$f = \frac{8}{\frac{\rho u_{avg} \lambda_D}{\eta} \cdot \frac{D_h}{D_h}} = \frac{8}{Re_{D_H}} \cdot \left(\frac{D_h}{\lambda_D}\right). \quad (12)$$

In the event that both an applied voltage and pressure drive the flow, the expression for friction factor is more involved and also depends on geometry. As an example consider fully developed laminar flow of a Newtonian fluid between two plates with the simultaneous application of an electric field in the flow direction. The application of the flow-direction Navier-Stokes equation with the inclusion of a body force term due to the applied electric field reduces to a linear differential equation. Thus the superposition principle allows the flow velocity to be expressed as the sum of EO only and pressure-driven only components:

$$u(y) = u_{EO} + u_{PD}. \quad (13)$$

Equation (8) for combined flow is therefore

$$f = \frac{4\tau_w}{\frac{1}{2}\rho(u_{EO,avg} + u_{PD,avg})^2} \quad (14)$$

The velocity distribution for the EO component $u_{EO}$ is given by (3) and the pressure-driven component $u_{PD}$ is the well-known parabolic velocity profile for fully-developed laminar flow given by

$$u_{PD}(y) = 6u_{PD,avg}\left[\frac{y}{a} - \left(\frac{y}{a}\right)^2\right], \quad (15)$$

where $a$ is the plate separation. Using (9) along with these expressions for velocity to find the wall shear stress, it can be shown after much manipulation that

$$f = \frac{96}{Re_{D_H}}\left(\frac{1+G}{1+12G\left(\frac{\lambda_D}{D_h}\right)}\right), \quad (16)$$

where $G$ represents the relative strength of the applied electric field to the applied pressure gradient and is given by

$$G \equiv \frac{\rho_e(\phi_2 - \phi_1)}{P_2 - P_1}. \quad (17)$$

Equations (16) and (17) assume a linearly applied pressure gradient and electric field and make use of the average velocity for Hagen-Poiseuille flow between two plates,

$$u_{PD,avg} = -\frac{dP}{dx}\frac{a^2}{12\eta}. \quad (18)$$

In the case of no applied electric field, $G \to 0$ and $f$ reduces to the familiar expression for pressure driven Hagen-Poiseuille flow between two plates given by

$$f = \frac{96}{Re_{D_H}} \quad (19).$$

For pure EO flow, $P_1 - P_2 = 0$ and $G \to \infty$. In this case $f$ reduces to (12).

Given that $f$ is independent of geometry for pure EO flow, (16) and (19) suggest that all combined EO and pressure-driven flows have friction factors of the form

$$f = \frac{C}{Re_{D_H}}\left(\frac{1+G}{1+(C/8)G\left(\frac{\lambda_D}{D_H}\right)}\right), \quad (20)$$

where $C$ is the appropriate coefficient for fully-developed laminar flow for the given geometry. Table 1 gives the values of $C$ for various geometries.

Table 1: Friction factors of fully developed laminar flow (Adapted from [9].)

| Cross Section | b/a | $C = f \cdot Re_{D_h}$ |
|---|---|---|
|  | - | 64 |
| 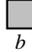 | 1.0 | 57 |
| 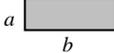 | 2.0 | 62 |
| 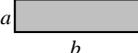 | 4.0 | 73 |
| 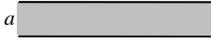 | ∞ | 96 |
|  | - | 53 |

## CONCLUSIONS

In this paper we show how to further modify the modified Bernoulli equation to incorporate electro-osmotic flow by adding electrical potential terms. In so doing we show that the friction factor for pure electro-osmotic is independent of geometry and inversely proportional to the Reynolds number based on the Debye length. Combined pressure-driven and electro-osmotic flows are also investigated and friction factor relationships developed. For combined flow the friction factor is a function of Reynolds number, the geometry of the flow channel, and the relative strength of the applied electric field to the pressure gradient. In the case of zero pressure gradient or zero applied electric field, friction factor expressions for pure electro-osmotic flow and laminar fully-developed pressure-driven flow are retrieved, respectively. The presentation of elements of electro-osmotic flow within a more traditional fluid mechanics framework should make the field of microfluidics more accessible to a general audience, as well as aid in the design and fabrication of microfluidic networks.

## ACKNOWLEDGMENTS

The authors wish to thank Dr. D.L. Morris in the Department of Chemistry at Rose-Hulman Institute of Technology for his generous collaborations in regard to this work.